\documentclass[a4paper]{jpconf}
\usepackage{graphicx}
\usepackage{amsmath,amssymb,amsopn,bm}
\newcommand{\beq}{\begin{equation}}
 \newcommand{\eeq}{\end{equation}}
 \newcommand{\beqa}{\begin{eqnarray}}
 \newcommand{\eeqa}{\end{eqnarray}}
 \DeclareMathOperator{\im}{Im}
 
\begin{document}
\title{Low mass dimuons within a hybrid approach}

\author{Elvira Santini$^1$, Marcus Bleicher$^2$}

\address{$^1$Institut f\"ur Theoretische Physik, Goethe-Universit\"at,
 Max-von-Laue-Str.~1, D-60438 Frankfurt am Main, Germany}

\address{$^2$Frankfurt Institute for Advanced Studies (FIAS), Ruth-Moufang-Str.~1,D-60438 Frankfurt am Main, Germany}

\ead{santini@th.physik.uni-frankfurt.de}

\begin{abstract}
We analyse dilepton emission from hot and dense hadronic matter 
using a hybrid approach based on the 
Ultrarelativistic Quantum Molecular Dynamics
(UrQMD) transport model with an intermediate hydrodynamic stage
for the description of heavy-ion collisions at relativistic energies.
Focusing on the enhancement with respect to the contribution from long-lived 
hadron decays after freeze-out observed at the SPS in the low mass region of 
the dilepton spectra (often referred to as ``the excess''), 
the relative importance of the emission from the 
equilibrium and the non-equilibrium stages is discussed.
\end{abstract}

\section{Introduction}
Electromagnetic probes, such as photons and lepton pairs, 
are penetrating probes of the hot and dense matter. 
Once created these particles pass the collision zone essentially 
without further interaction and can therefore mediate valuable information 
on the electromagnetic response of the strongly interacting medium.
The analysis of such response 
is tightly connected to the investigation of the in-medium modification 
of the vector meson properties. Vector mesons can directly decay into a 
lepton-antilepton pair. One therefore aims to infer information on the 
modifications induced by the medium on specific properties of the vector 
meson, such as its mass and/or its width, from the invariant mass dilepton 
spectra. 

Measurements of dileptons in heavy ion collisions have indeed revealed 
many exciting phenomena, triggering a copious theoretical activity. 
A first generation of ultra-relativistic heavy ion collision experiments 
performed in the nineties observed an enhancement of dilepton production 
in heavy systems at low invariant mass as
compared to conventional hadronic cocktails and models
\cite{Agakishiev:1995xb,Mazzoni:1994rb}.
The enhancement could be later explained by the inclusion 
of a substantial medium modification of the $\rho$ meson properties, see e.g. 
Refs.~\cite{Rapp:1999ej,Gale:2003iz,Rapp:2009yu} for reviews 
or the review on photons and dileptons by S.~Bathe in this volume.
However, no discrimination between a ``melting'' of the $\rho$ meson 
spectral function as expected within many-body hadronic models
\cite{Rapp:1997fs,Friman:1997tc,Peters:1997va,Lutz:2001mi,Santini:2008pk} 
and a dropping 
of the $\rho$ meson mass
according to the Brown-Rho scaling hypothesis \cite{Brown:1991kk}
and the Hatsuda and Lee sum rule prediction
\cite{Hatsuda:1991ez} could be achieved.
Recently, a substantial improvement in statistics and mass resolution in 
low-mass dilepton spectra has been achieved by the NA60 collaboration 
\cite{Arnaldi:2006jq}, who measured dimuon production in In-In collisions 
at 158 A GeV. The improved experimental accuracy enabled a subtraction of 
final-state hadron decay contributions (the so-called cocktail), 
except for $\rho$ and charm decays. 
The dimuon spectra of the remaining ``excess'' over the cocktail strongly
favour the broadening over the dropping mass scenario.
Typically, model calculations aimed at the interpretation of the 
data require a convolution of the dilepton emission rates over a realistic 
space-time model of the heavy-ion collision. Assuming local equilibrium, thermal 
fireball/hydrodynamics calculations have been performed over the last years and quite successfully 
applied to the interpretation of the NA60 data 
(see e.g. Fig.4 in Ref.~\cite{Arnaldi:2008fw}).
Here, we present recent results on dimuon emission 
obtained using a hybrid approach for the description of the 
evolution dynamics. In such an approach, both thermal and 
non-thermal sources play a noticeable role and can be separately investigated. 
We will focus 
on the low invariant mass region of the dilepton spectra ($M<1$ GeV), where 
the $\rho^*\rightarrow \mu\mu$ emission plays the dominant role. 
Extension of the model to the intermediate mass region ($1.0<M<1.5$ GeV) 
requires addition of further contributions and is 
currently under development.

\section{Dilepton phenomenology within a hybrid model}

The dynamics
of the In+In collisions is simulated employing a hybrid approach 
based on the integration of an ideal hydrodynamic evolution into the 
UrQMD transport model  \cite{Petersen:2008dd}.
This approach has 
been successfully 
applied to many bulk observables at SPS energies 
\cite{Petersen:2008dd,Petersen:2009vx,Petersen:2009mz,Petersen:2009zi,Graf:2010pm} and first applications 
to electromagnetic probes have been performed recently 
\cite{Bauchle:2009ep,Santini:2009nd}. Here, we describe it only briefly and refer the reader 
to the indicated references for details.

During the first stage of the evolution the particles are described as a 
purely hadronic cascade within UrQMD. The coupling to the hydrodynamical 
evolution proceeds when the two Lorentz-contracted nuclei have 
passed through each other.
At this time, the spectators continue to propagate 
in the cascade and all other hadrons
are mapped to the hydrodynamic grid.  
Subsequently, a $(3+1)$ ideal hydrodynamic evolution is performed using
the SHASTA algorithm \cite{Rischke:1995ir,Rischke:1995mt}. The hydrodynamic
evolution is gradually merged into the hadronic cascade. 
Transverse slices, of thickness 0.2 fm, are transformed to particles 
whenever in all
cells of each individual slice the energy density  drops below
five times the ground state energy density. 
The employment of such a gradual transition allows us to 
obtain a rapidity independent transition temperature 
without artificial time dilatation effects.
When merging, the hydrodynamic fields are transformed to particle
degrees of freedom via the Cooper-Frye equation. 
The created particles proceed in their evolution in
the hadronic cascade where final state interactions and decays of the particles occur within the UrQMD framework.

Concerning dilepton emission, during the locally equilibrated 
hydrodynamic stage the production of lepton pairs is described by
radiation rates for a strongly interacting medium in thermal equilibrium. 
Invoking vector 
meson dominance the latter can be related, at low invariant masses, 
to the spectral properties of the 
vector mesons, 
with the $\rho$ meson giving the dominant contribution. 
The thermal dilepton rate reads then \cite{Rapp:1999ej}:
\beq
\frac{d^8 N_{ll}}{d^4 x d^4 q}=-\frac{\alpha^2m_\rho^4}{\pi^3 g_\rho^2}\frac{L(M^2)}{M^2}f_B(q_0;T)
\im D_\rho(M,q;T,\mu_B) \, ,  
\label{rate}
\eeq
where $\alpha$ denotes the fine structure constant, 
$M^2=q_0^2-q^2$ the dilepton invariant mass squared, $f_B$ the Bose 
distribution function 
(for a moving fluid this must be substituted with the J\"{u}ttner function), and 
$L(M)$ a lepton phase space factor that quickly approaches one above the lepton pair threshold. 
The electromagnetic response of the strongly interacting medium is then contained in $\im D_\rho(M,q;T,\mu_B)$, 
the imaginary part of the in-medium $\rho$ meson propagator,
\beq
D_\rho(M,q;T,\mu_B)=\frac{1}{M^2-m_\rho^2-\Sigma_\rho(M,q;T,\mu_B)}.
\label{rhoprop}
\eeq
In this application, the self-energy
contributions taken into account are $\Sigma_\rho=\Sigma^0+\Sigma^{\rho\pi}+\Sigma^{\rho N}$,
where $\Sigma^0$ is the vacuum self-energy and $\Sigma^{\rho\pi}$ and $\Sigma^{\rho N}$
denote the contribution to the self-energy due to the direct
interactions of the $\rho$ with, respectively, pions and nucleons of the
surrounding heat bath. The self-energies have been 
calculated according to Ref. \cite{Eletsky:2001bb}, where they were 
evaluated in terms of empirical 
scattering amplitudes from resonance dominance at low energies and 
Regge-type behaviour at high energy. 
Finally, in  the evolution stage that precedes or follows the 
hydrodynamical phase dimuon emission from the 
$\rho$-meson is calculated  employing the time integration method  
that has long been 
applied in the transport description of dilepton emission 
(see e.g. \cite{Schmidt:2008hm}).

\begin{figure}
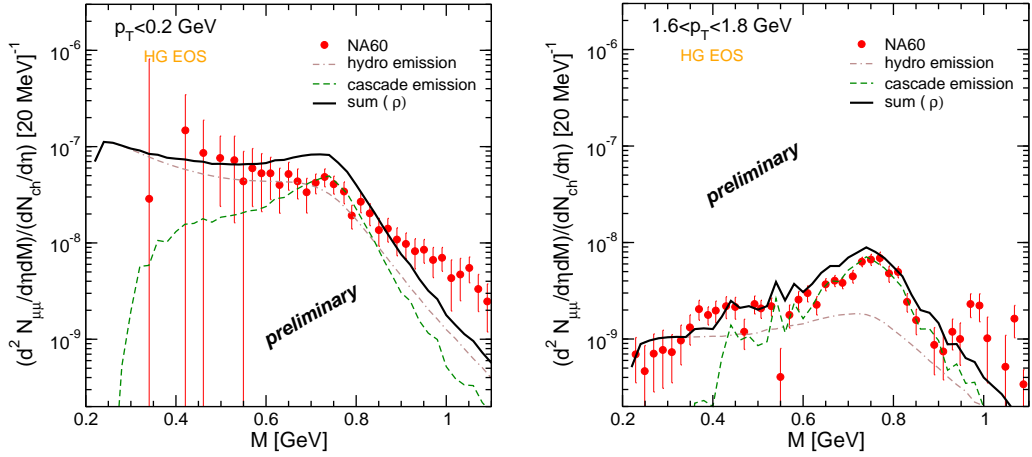

  \begin{center}
  \includegraphics[width=0.4\textwidth]{proc.HGim.am.pt.le.0.2.eps}
\hspace{.4cm}
  \includegraphics[width=0.4\textwidth]{proc.HGim.am.1.6.pt.1.8.eps}
  \end{center}
  \caption{\label{na_60_HGim.am.pt.le.0.2} Left panel: Acceptance-corrected invariant 
 mass spectra of the excess dimuons in In-In collisions at 158$A$ 
 GeV for transverse pair momenta $p_T<0.2$ GeV, 
 compared to hybrid model calculations based on thermal radiation from 
 in-medium modified $\rho$ meson spectral function and non-thermal cascade 
 emission. Experimental data from Ref.~\cite{Arnaldi:2008fw}. Right panel: Same as in the left panel, but for the transverse momenta window  $1.6<p_T<1.8$ GeV.}
\end{figure}

In Fig.~\ref{na_60_HGim.am.pt.le.0.2} 
hybrid model calculations are compared to recent acceptance-corrected 
NA60 data 
\cite{Arnaldi:2008fw}. 
The calculations have been exemplary performed using a hadron gas equation 
of state (HG-EoS) for the hydrodynamical evolution.
Since in such a scenario all the energy density is stored in hadronic 
degrees of freedom, for the sake of consistency 
dimuon emission according to Eq.~(\ref{rate}) has been 
evaluated during the whole hydrodynamical phase, 
hottest cells included 
(in the central region some cells have 
initial temperatures up to $T\sim$240 MeV).
In general, however, if a phase transition from quark-gluon 
to hadronic matter occurs, the hadronic thermal
rate would be only a fraction of the total thermal rate.
Keeping in mind that the total life-time of the fireball affects 
the total dilepton yield too, we can quite generally say that 
the inclusion of a phase in which quarks and gluons 
are the relevant degrees of freedom would result in a reduction 
of the hadronic emission with respect to the calculation presented here, 
provided that the life-time of the fireball is comparable for 
the two equations of state.
Calculations with such an EoS are on-going and suggest 
this reduction 
to amount to about 30\%-40\% across the entire 
dimuon invariant mass region considered here. Anyway, 
this baseline calculation  
performed with the HG-EoS still allows
to  point out qualitatively the 
main features of the present approach. 
More systematic studies and discussions will be 
presented elsewhere \cite{Elviranext} in the next future.

We observe that the cascade emission dominates the invariant mass region 
around the vector meson peak 
for both low and intermediate transverse pair momenta $p_T$. 
At low $p_T$ 
(left panel of Fig.~\ref{na_60_HGim.am.pt.le.0.2}), 
the  very low invariant masses, $M<0.5$ GeV, 
are filled by the thermal 
radiation with in-medium spectral function. 
The sum of both contributions, however, 
leads to an overestimation of the vector meson 
peak region at low transverse momenta of the dilepton pair. 
The reason for the discrepancy might partially lie in the specific 
spectral function used here 
and/or, presumably more severely, on the eventual presence of not yet negligible residual 
in-medium modification 
of the $\rho$ meson spectral function during the cascade stage, 
that are here neglected. An additional source of uncertainty 
is the dependence of the total thermal yield on the specific criterium adopted 
to perform the switch from the hydrodynamic 
to the transport description. 
Further investigations are required to clarify these model dependences 
and quantify their effect on the total dilepton emission.
With increasing  $p_T$ (right panel of Fig.~\ref{na_60_HGim.am.pt.le.0.2}), 
dilepton emission at very low invariant masses 
is reduced and the total spectra are almost completely determined by 
the cascade emission.  This is not trivial. 
Indeed, thermal emission had already shown 
discrepancies for $p_T>1$  GeV \cite{vanHees:2007th}, pointing to the 
necessity to account for non-thermal contributions. In the present approach, 
the latters appear quite naturally.

\section{Acknowledgments}
The authors thanks the organizers for the exciting and sunny conference.
This work was supported by the Hessen Initiative 
for Excellence (LOEWE) through the Helmholtz International Center for FAIR 
(HIC for FAIR). We thank the Center for Scientific Computing for providing 
computational resources.

\section*{References}

\bibliography{/home/santini/paper_frank/paper_na60/biblio}

\providecommand{\newblock}{}
\begin{thebibliography}{10}
\expandafter\ifx\csname url\endcsname\relax
  \def\url#1{{\tt #1}}\fi
\expandafter\ifx\csname urlprefix\endcsname\relax\def\urlprefix{URL }\fi
\providecommand{\eprint}[2][]{\url{#2}}

\bibitem{Agakishiev:1995xb}
Agakishiev G {\em et~al.\/} (CERES) 1995 {\em Phys. Rev. Lett.\/} {\bf 75}
  1272--1275

\bibitem{Mazzoni:1994rb}
Mazzoni M~A (HELIOS/3.) 1994 {\em Nucl. Phys.\/} {\bf A566} 95c--102c

\bibitem{Rapp:1999ej}
Rapp R and Wambach J 2000 {\em Adv. Nucl. Phys.\/} {\bf 25} 1
  (\textit{Preprint} \eprint{hep-ph/9909229})

\bibitem{Gale:2003iz}
Gale C and Haglin K~L 2003 {Electromagnetic radiation from relativistic nuclear
  collisions} {\em {Quark-gluon plasma. Vol. 3}\/} ed Hwa R~C and Wang X~N
  (River Edge, USA: World Scientific) p 348 (\textit{Preprint}
  \eprint{hep-ph/0306098})

\bibitem{Rapp:2009yu}
Rapp R, Wambach J and van Hees H 2009 {The Chiral Restoration Transition of QCD
  and Low Mass Dileptons} {\em {Relativistic Heavy Ion Physics}\/} ed Stock R
  (Landolt-Bernstein series) (\textit{Preprint} \eprint{0901.3289})

\bibitem{Rapp:1997fs}
Rapp R, Chanfray G and Wambach J 1997 {\em Nucl. Phys.\/} {\bf A617} 472--495
  (\textit{Preprint} \eprint{hep-ph/9702210})

\bibitem{Friman:1997tc}
Friman B and Pirner H~J 1997 {\em Nucl. Phys.\/} {\bf A617} 496--509
  (\textit{Preprint} \eprint{nucl-th/9701016})

\bibitem{Peters:1997va}
Peters W, Post M, Lenske H, Leupold S and Mosel U 1998 {\em Nucl. Phys.\/} {\bf
  A632} 109--127 (\textit{Preprint} \eprint{nucl-th/9708004})

\bibitem{Lutz:2001mi}
Lutz M~F~M, Wolf G and Friman B 2002 {\em Nucl. Phys.\/} {\bf A706} 431--496
  (\textit{Preprint} \eprint{nucl-th/0112052})

\bibitem{Santini:2008pk}
Santini E {\em et~al.\/} 2008 {\em Phys. Rev.\/} {\bf C78} 034910
  (\textit{Preprint} \eprint{0804.3702})

\bibitem{Brown:1991kk}
Brown G~E and Rho M 1991 {\em Phys. Rev. Lett.\/} {\bf 66} 2720--2723

\bibitem{Hatsuda:1991ez}
Hatsuda T and Lee S~H 1992 {\em Phys. Rev.\/} {\bf C46} 34--38

\bibitem{Arnaldi:2006jq}
Arnaldi R {\em et~al.\/} (NA60) 2006 {\em Phys. Rev. Lett.\/} {\bf 96} 162302
  (\textit{Preprint} \eprint{nucl-ex/0605007})

\bibitem{Arnaldi:2008fw}
Arnaldi R {\em et~al.\/} (NA60) 2009 {\em Eur. Phys. J.\/} {\bf C61} 711--720
  (\textit{Preprint} \eprint{0812.3053})

\bibitem{Petersen:2008dd}
Petersen H, Steinheimer J, Burau G, Bleicher M and Stocker H 2008 {\em Phys.
  Rev.\/} {\bf C78} 044901 (\textit{Preprint} \eprint{0806.1695})

\bibitem{Petersen:2009vx}
Petersen H and Bleicher M 2009 {\em Phys. Rev.\/} {\bf C79} 054904
  (\textit{Preprint} \eprint{0901.3821})

\bibitem{Petersen:2009mz}
Petersen H, Steinheimer J, Bleicher M and Stocker H 2009 {\em J. Phys.\/} {\bf
  G36} 055104 (\textit{Preprint} \eprint{0902.4866})

\bibitem{Petersen:2009zi}
Petersen H, Mitrovski M, Schuster T and Bleicher M 2009 {\em Phys. Rev.\/} {\bf
  C80} 054910 (\textit{Preprint} \eprint{0903.0396})

\bibitem{Graf:2010pm}
Graf G {\em et~al.\/} 2010 {\em J. Phys.\/} {\bf G37} 094010 (\textit{Preprint}
  \eprint{1001.4937})

\bibitem{Bauchle:2009ep}
Baeuchle B and Bleicher M 2010 {\em Phys. Rev.\/} {\bf C81} 044904
  (\textit{Preprint} \eprint{0905.4678})

\bibitem{Santini:2009nd}
Santini E, Petersen H and Bleicher M 2010 {\em Phys. Lett.\/} {\bf B687}
  320--326 (\textit{Preprint} \eprint{0909.4657})

\bibitem{Rischke:1995ir}
Rischke D~H, Bernard S and Maruhn J~A 1995 {\em Nucl. Phys.\/} {\bf A595}
  346--382 (\textit{Preprint} \eprint{nucl-th/9504018})

\bibitem{Rischke:1995mt}
Rischke D~H, Pursun Y and Maruhn J~A 1995 {\em Nucl. Phys.\/} {\bf A595}
  383--408 (\textit{Preprint} \eprint{nucl-th/9504021})

\bibitem{Eletsky:2001bb}
Eletsky V~L, Belkacem M, Ellis P~J and Kapusta J~I 2001 {\em Phys. Rev.\/} {\bf
  C64} 035202 (\textit{Preprint} \eprint{nucl-th/0104029})

\bibitem{Schmidt:2008hm}
Schmidt K {\em et~al.\/} 2009 {\em Phys. Rev.\/} {\bf C79} 064908
  (\textit{Preprint} \eprint{0811.4073})

\bibitem{Elviranext}
Santini E {\em et~al.\/} {} in preparation

\bibitem{vanHees:2007th}
van Hees H and Rapp R 2008 {\em Nucl. Phys.\/} {\bf A806} 339
  (\textit{Preprint} \eprint{0711.3444})

\end{thebibliography}

\end{document}